\documentclass[a4,11pt]{article}
\usepackage{graphicx} 
\usepackage{amsmath}
\usepackage{amsfonts}
\usepackage{latexsym}
\usepackage{amssymb}
\makeatletter
 
  \@addtoreset{equation}{section}
 \makeatother

\begin{document}
\title{Perturbative Expansion Technique for Non-linear FBSDEs with Interacting Particle Method~\footnote{
This research is supported by CARF (Center for Advanced Research in Finance) and 
the global COE program ``The research and training center for new development in mathematics.''
All the contents expressed in this research are solely those of the authors and do not represent any views or 
opinions of any institutions. 
The authors are not responsible or liable in any manner for any losses and/or damages caused by the use of any contents in this research.
}
}

\author{Masaaki Fujii\footnote{Graduate School of Economics, The University of Tokyo},
Akihiko Takahashi\footnote{Graduate School of Economics, The University of Tokyo}
}
\date{
First version: April 12, 2012\\ 
Current version: April 23, 2012
}
\maketitle



\newtheorem{definition}{Definition}
\newtheorem{assumption}{$[$ A}
\newtheorem{condition}{$[$ C}
\newtheorem{lemma}{Lemma}
\newtheorem{proposition}{Proposition}
\newtheorem{theorem}{Theorem}
\newtheorem{remark}{Remark}
\newtheorem{example}{Example}
\newtheorem{corollary}{Corollary}
\def\n{{\bf n}}
\def\A{{\bf A}}
\def\B{{\bf B}}
\def\C{{\bf C}}
\def\D{{\bf D}}
\def\E{{\bf E}}
\def\F{{\bf F}}
\def\G{{\bf G}}
\def\H{{\bf H}}
\def\I{{\bf I}}
\def\J{{\bf J}}
\def\K{{\bf K}}
\def\L{{\bf L}}
\def\M{{\bf M}}
\def\N{{\bf N}}
\def\O{{\bf O}}
\def\P{{\bf P}}
\def\Q{{\bf Q}}
\def\R{{\bf R}}
\def\S{{\bf S}}
\def\T{{\bf T}}
\def\U{{\bf U}}
\def\V{{\bf V}}
\def\W{{\bf W}}
\def\X{{\bf X}}
\def\Y{{\bf Y}}
\def\Z{{\bf Z}}
\def\cala{{\cal A}}
\def\calb{{\cal B}}
\def\calc{{\cal C}}
\def\cald{{\cal D}}
\def\cale{{\cal E}}
\def\calf{{\cal F}}
\def\calg{{\cal G}}
\def\calh{{\cal H}}
\def\cali{{\cal I}}
\def\calj{{\cal J}}
\def\calk{{\cal K}}
\def\call{{\cal L}}
\def\calm{{\cal M}}
\def\caln{{\cal N}}
\def\calo{{\cal O}}
\def\calp{{\cal P}}
\def\calq{{\cal Q}}
\def\calr{{\cal R}}
\def\cals{{\cal S}}
\def\calt{{\cal T}}
\def\calu{{\cal U}}
\def\calv{{\cal V}}
\def\calw{{\cal W}}
\def\calx{{\cal X}}
\def\caly{{\cal Y}}
\def\calz{{\cal Z}}
%
\def\sskip{\hspace{0.5cm}}
\def\simleq{ \raisebox{-.7ex}{\em $\stackrel{{\textstyle <}}{\sim}$} }
\def\leqsim{ \raisebox{-.7ex}{\em $\stackrel{{\textstyle <}}{\sim}$} }
\def\ep{\epsilon}
\def\half{\frac{1}{2}}
\def\iku{\rightarrow}
\def\Iku{\Rightarrow}
\def\ikup{\rightarrow^{p}}
\def\inclusion{\hookrightarrow}
\def\cadlag{c\`adl\`ag\ }
\def\up{\uparrow}
\def\down{\downarrow}
\def\doti{\Leftrightarrow}
\def\douti{\Leftrightarrow}
\def\dochi{\Leftrightarrow}
\def\douchi{\Leftrightarrow}%
\def\yy{\\ && \nonumber \\}
\def\y{\vspace*{3mm}\\}
\def\nn{\nonumber}
\def\be{\begin{equation}}
\def\ee{\end{equation}}
\def\bea{\begin{eqnarray}}
\def\eea{\end{eqnarray}}
\def\beas{\begin{eqnarray*}}
\def\eeas{\end{eqnarray*}}
%
\def\hd{\hat{D}}
\def\hv{\hat{V}}
\def\hsd{{\hat{d}}}
\def\hx{\hat{X}}
\def\hsx{\hat{x}}
\def\bsx{\bar{x}}
\def\bsd{{\bar{d}}}
\def\bx{\bar{X}}
\def\ba{\bar{A}}
\def\bb{\bar{B}}
\def\bc{\bar{C}}
\def\bv{\bar{V}}
\def\balpha{\bar{\alpha}}
\def\bbalpha{\bar{\bar{\alpha}}}
\def\combi{\l(\begin{array}{c}\alpha\\ \beta \end{array}\r)}
\def\f{^{(1)}}
\def\s{^{(2)}}
\def\ss{^{(2)*}}
\def\l{\left}
\def\r{\right}
\def\a{\alpha}
\def\b{\beta}
\def\L{\Lambda}

\def\E{{\bf E}}
\def\P{{\bf P}}
\def\Q{{\bf Q}}
\def\R{{\bf R}}

\def\calf{{\cal F}}
\def\calp{{\cal P}}
\def\calq{{\cal Q}}

\def\ep{\epsilon}
\def\del{\delta}
\def\al{\alpha}
\def\part{\partial}
\def\ol{\overline}

\def\cdott{\cdot\cdot}

\def\yy{\\ && \nonumber \\}
\def\y{\vspace*{3mm}\\}
\def\nn{\nonumber}
\def\be{\begin{equation}}
\def\ee{\end{equation}}
\def\bea{\begin{eqnarray}}
\def\eea{\end{eqnarray}}
\def\beas{\begin{eqnarray*}}
\def\eeas{\end{eqnarray*}}
\def\l{\left}
\def\r{\right}
\vspace{10mm}

\begin{abstract}
In this paper, we propose an efficient Monte Carlo implementation of non-linear FBSDEs as a system of 
interacting particles inspired by the ideas of branching diffusion method.
It will be particularly useful to investigate large and complex systems, and hence it is a 
good complement of our previous work presenting an analytical perturbation procedure 
for generic non-linear FBSDEs.
There appear multiple species of particles, where the first 
one follows the diffusion of the original underlying state,
and the others the Malliavin derivatives with a grading structure.
The number of branching points are capped by the order of perturbation, which 
is expected to make the scheme less numerically intensive.
The proposed method can be applied to semi-linear problems, such as American and Bermudan options,
Credit Value Adjustment (CVA), and even fully non-linear issues, such as the optimal portfolio problems in 
incomplete and/or constrained markets, feedbacks from large investors,  
and also the analysis of various risk measures.

\end{abstract}
\vspace{17mm}
{\bf Keywords :}
BSDE, FBSDE, Asymptotic Expansion, Malliavin Derivative, interacting particle method,
branching diffusion

\newpage
\section{Introduction}
The forward backward stochastic differential equations (FBSDEs) were first introduced by Bismut (1973)~\cite{Bismut}, 
and then later extended by Pardoux and Peng (1990)~\cite{P-Peng} for general non-linear cases.
They were found particularly relevant for optimal portfolio and indifference pricing issues~
in incomplete and/or constrained markets. Their financial applications are discussed in 
details in, for example, El~Karoui, Peng and Quenez (1997a)~\cite{ElKaroui},
Ma and Yong (2000)~\cite{Ma} and a recent book edited by Carmona (2009)~\cite{Carmona}.
The importance of FBSDEs will increase in coming years even among practitioners where the new 
financial regulations will put significant constraints on available assets and trading strategies.

In the recent paper, Fujii \& Takahashi (2011)~\cite{analytic_FBSDE} proposed a new 
perturbative solution technique for generic non-linear FBSDEs. 
It was shown that a non-linear FBSDE can be decomposed into a series of 
linear and decoupled FBSDEs by treating a non-linear driver and feedback terms
as perturbations to the corresponding decoupled free system. 
In particular, it allows analytical explicit expressions for the backward components with the 
help of the asymptotic expansion technique~(See, for example~\cite{T,KT,TY,asymptotic3}.). 
A backward component of the diffusion part was shown to be obtained
by directly considering dynamics of the stochastic flow, which denotes a Malliavin derivative of
the underling state process, or simply applying It\^o formula to the result of the other part.
In Fujii \& Takahashi (2012)~\cite{qbFBSDE}, the method was applied to a quadratic-growth 
FBSDE appearing in an incomplete financial market with stochastic volatility. Explicit 
expressions for both of the backward components were obtained up to the third order of the 
volatility of volatility. The comparison 
to the exact solution with Cole-Hopf transformation demonstrated effectiveness of the 
perturbative expansion. 

Notice the fact that one can already apply standard Monte Carlo simulation to the results
obtained in each order of the perturbative expansion in \cite{analytic_FBSDE}.
However, due to its convoluted nature, it contains multi-dimensional time integrations of 
expectation values which make the naive applications too time consuming,  particularly for the evaluation of 
higher order perturbation terms.
To handle this problem, we applied the idea of particle representation 
used in branching diffusion models, such as in McKean (1975)~\cite{McKean}.
There, the convoluted expectation is compressed into a single standard expectation by introducing 
an intensity of the particle interaction. 
McKean~\cite{McKean} applied the method to solve a particular type of semi-linear PDE, where a single particle 
splits into two at each interaction time and creates a cascade of the identical particles.
Note that, our method is not directly related to McKean~\cite{McKean} since the interested system is already 
decomposed into a set of linear problems, although we have used the similar 
particle representation to avoid nested simulations.

The analysis of branching Markov process and related problems in semi-linear PDEs has a long history. 
Some of the well-known works are Fujita (1966)~\cite{fujita}, Ikeda, Nagasawa~\&~Watanabe (1965,1966,1968)~\cite{inw1,inw2,inw3},
Ikeda et.al.~(1996,1997)~\cite{Seminar} and Nagasawa~\&~Sirao (1969)~\cite{Nagasawa}.
As for a recent work, in particular,  Chakraborty~\&~L\'opez-Mimbela (2008) used particle 
representation where the number of offspring at each interaction point
is randomly drawn by some probability distribution, which can be finitely many or infinite.
The authors used the branching particle representation~\footnote{The same branching representation is
already seen in \cite{inw3}, for example.} to study 
the existence of global solutions for semi-linear PDEs with a non-linear driver
given by a generic polynomial function~\footnote{For recent developments and reviews of the particle 
methods, see for examples ~\cite{DelMoral,Filter}. There exist a significant amount of 
works related to branching diffusion in 1960's and 70's. There are also a vast range of new applications 
and enhancements in biology, such as gene mutation and population growth problems, as well as
in engineering issues. We have not yet obtained 
the whole picture of research history related to branching diffusion and are welcoming information 
from those familiar with the topic.}.
Recently, Henry-Labord\`ere (2012)~\cite{Labordere} introduced a particle representation 
to study the semi-linear problems in finance. He called it marked branching diffusion 
and has discussed its application to efficiently calculate CVA (credit value adjustment) in one-shot Monte Carlo simulation. 
He also referred to its application to other semi-linear problems, 
such as American options, as well as its possible extension to truly non-linear problems by using Malliavin 
derivatives.

In the current paper, we combine the idea of particle representation and the perturbation technique
developed in the previous work~\cite{analytic_FBSDE}.
We provide a straightforward simulation scheme to solve fully-nonlinear decoupled as well as 
coupled FBSDEs at each order of perturbative approximation.
In contrast to the direct application of branching diffusion method,
the number of branching points are capped by the order of perturbative expansion,
which is due to the linearity of the decomposed FBSDE system. This property 
is expected to make Monte Carlo simulation less numerically intensive.
Our method can be applied to semi-linear problems, such as American and Bermudan options~\footnote{A BSDE formulation for an American option was shown in El~Karoui etal. (1997b)~\cite{ElKaroui_2}, which was 
recently studied by Labart~\&~Lelong (2011)~\cite{Labart} based on regression based Monte Carlo simulation.},
Credit Value Adjustment (CVA) as special examples. It can be also applied to 
fully non-linear (and fully coupled) issues, such as the optimal portfolio 
problems in incomplete and/or constrained market, analysis for various risk measures 
as well as for the feedbacks from so-called large investors. 
Concrete applications of the new method will be published separately~\cite{BSDEAmerican}.

\section{Setup}
We first consider generic decoupled non-linear FBSDEs.
Let us use the same setup assumed in the work~\cite{analytic_FBSDE}.
The probability space is taken as $(\Omega,\calf, P)$ and $T\in(0,\infty)$ denotes
some fixed time horizon. $W_t=(W_t^1,\cdots,W_t^r)^*$, $0\leq t\leq T$ is $\mathbb{R}^r$-valued
Brownian motion defined on $(\Omega,\calf,P)$, and $(\calf_t)_{\{0\leq t \leq T\}}$ stands for 
P-augmented natural filtration generated by the Brownian motion.

We consider the following forward-backward stochastic differential equation (FBSDE)
\bea
dV_s&=&-f(X_s,V_s,Z_s)ds+Z_s\cdot dW_s \\
V_T&=&\Psi(X_T)
\eea
where $V$ takes the value in $\mathbb{R}$, and $X_t\in \mathbb{R}^d$ is assumed to follow a generic Markovian forward SDE 
\be
dX_s=\gamma_0(X_s)ds+\gamma(X_s)\cdot dW_s ~.
\label{XSDE}
\ee
Here, we absorbed an explicit dependence on time to $X$ by allowing some of its components  
can be a time itself. $\Psi(X_T)$ denotes the terminal payoff where 
$\Psi(x)$ is a deterministic function of $x$.
The following approximation procedures can be applied in the same way 
also in the presence of coupon payments. 
$Z$ and $\gamma$ take values in $\mathbb{R}^r$ and $\mathbb{R}^{d\times r}$ respectively, 
and "$\cdot$" in front of the $dW$ represents the summation for the 
components of $r$-dimensional Brownian motion.
Throughout this paper, we are going to assume that the appropriate regularity conditions are satisfied 
for the necessary treatments.

Let us fix the initial time as $t$. We denote the Malliavin derivative of $X_u~(u\geq t)$ at time $t$
as
\bea
\cald_t X_u \in \mathbb{R}^{r\times d}
\eea 
Its dynamics in terms of the future time $u$ is specified by the well-known stochastic flow:
\bea
d(Y_{t,u})^i_j&=&\part_k \gamma_0^i(X_u)(Y_{tu})^k_j du+\part_k \gamma_a^i(X_u)(Y_{tu})^k_j dW^a_u \nn \\
(Y_{t,t})^i_j&=&\del^i_j
\label{YSDE}
\eea
where $\part_k$ denotes the differential with respect to the k-th component of $X$, and $\del^i_j$ 
denotes Kronecker delta. Here, $i$ and $j$ run through $\{1,\cdots,d\}$ and $\{1,\cdots,r\}$ for $a$.
Throughout the paper, we adopt Einstein notation which assumes the summation of all the 
paired indexes.
Using the known chain rule of Malliavin derivative, one sees
\bea
(\cald_tX_u^i)=\int_t^u \part_k \gamma^i_0(X_s)(\cald_t X_s^k)ds+
\int_t^u \part_k \gamma^i(X_s)(\cald_t X_s^k)\cdot dW_s+\gamma^i(X_t)
\eea
and hence it satisfies
\bea
(\cald_t X_u^i)_{a}&=&(Y_{t,u})^i_j\gamma^j_a(X_t)=(Y_{t,u}\gamma(X_t))^i_a
\eea
where "$a$" is the index of $r$-dimensional Brownian motion.

\section{Expansion into a series of Linear FBSDE System}
\label{sec-expandedFBSDEs}
Following the perturbative method proposed in~\cite{analytic_FBSDE}, let us 
introduce the perturbation parameter $\ep$ and then write the equation as
\bea
\begin{cases}
& dV_s^{(\ep)}=-\ep f(X_s,V_s^{(\ep)},Z_s^{(\ep)})ds+Z_s^{(\ep)}\cdot dW_s \\
& V_T^{(\ep)}=\Psi(X_T)
\end{cases}
\eea
where $\ep=1$ corresponds to the original model~\footnote{It is possible to extract the 
linear term from the driver and treat separately. Here, we simply leave it in a driver, or 
work in a "discounted" base to remove linear term in $V$.}.
We suppose that the solution can be expanded in a power series of $\ep$:
\bea
V_t^{(\ep)}&=&V_t^{(0)}+\ep V_t^{(1)} + \ep^2 V_t^{(2)}+\ep^3 V_t^{(3)}+\cdots \\
Z_t^{(\ep)}&=&Z_t^{(0)}+\ep Z_t^{(1)} + \ep^2 Z_t^{(2)}+\ep^3 Z_t^{(3)}+\cdots 
\eea
If the non-linearity is sub-dominant, one can expect to obtain reasonable 
approximation of the original system by putting $\ep=1$ at the end of calculation.
\\

The dynamics of each pair $(V^{(i)},Z^{(i)})$ can be easily derived as follows:\\
{\bf{Zero-th order}}
\bea
\begin{cases}
& dV_s^{(0)}=Z_s^{(0)}\cdot dW_s \\
& V_T^{(0)}=\Psi(X_T) 
\end{cases}
\eea
{\bf{First order}}
\be
\begin{cases}
& dV_s^{(1)}=-f(X_s,V_s^{(0)},Z_s^{(0)})ds+Z_s^{(1)}\cdot dW_s\\
& V_T^{(1)}=0
\end{cases}
\ee
{\bf{Second order}}
\be
\begin{cases}
& dV_s^{(2)}=-\left\{V_s^{(1)}\frac{\part}{\part v}+
(Z_s^{a(1)})\frac{\part}{\part z^a}\right\}f(X_s,V_s^{(0)},Z_s^{(0)})ds+Z_s^{(2)}\cdot dW_s \\
& V_T^{(2)}=0
\label{bsde2nd}
\end{cases}
\ee
{\bf{Third order}}
\be
\begin{cases}
& dV_s^{(3)}=-\Bigl\{V_s^{(2)}\frac{\part}{\part v}+Z_s^{2(a)}\frac{\part}{\part z^a}+ \frac{1}{2}(V_s^{(1)})^2\frac{\part^2}{\part v^2}+V_s^{(1)}Z_s^{a(1)}\frac{\part^2}{\part v\part z^a} \\
&\hspace{15mm}+\frac{1}{2}Z_s^{a(1)}Z_s^{b(1)}\frac{\part^2}{\part z^a\part z^b}\Bigr\}f(X_s,V_s^{(0)},Z_s^{(0)})ds +Z_s^{(3)}\cdot dW_s\\ 
&V_T^{(3)}=0
\end{cases}
\ee
One can continue to an arbitrary higher order in the same way.
\begin{center}
$\cdots\cdots\cdots$
\end{center}
Note that the higher order backward components $(V^{(n)},Z^{(n)})_{\{n\geq 1\}}$ are always outside of 
the non-linear functions. This property arises naturally due to the very nature of perturbation.
As we shall see, this is crucial to suppress the number of particles in the numerical simulation.
\section{Interacting Particle Interpretation}
Let us fix the initial time $t$ and set $X_t=x_t$.
\subsection{$\ep$-0th Order}
For the zeroth order, it is easy to see
\bea
 V_t^{(0)}&=&\mathbb{E}\Bigl[\Psi(X_T)\Bigr|\calf_t\Bigr] \\
 Z_t^{a(0)}&=&\mathbb{E}\Bigl[\part_i \Psi(X_T)(\cald^a_t X_T^i)\Bigr|\calf_t\Bigr] \nn \\
&=&\mathbb{E}\Bigl[\part_i \Psi(X_T) (Y_{tT}\gamma(X_t))^{i}_a\Bigr|\calf_t \Bigr]
\eea
It is clear that they can be evaluated by standard Monte Carlo simulation.
However, for their use in higher order approximation, it is crucial to obtain 
explicit approximate expressions for these two quantities.
As proposed in \cite{analytic_FBSDE}, we use asymptotic expansion technique~\cite{T,KT,TY, asymptotic3}
for this purpose. When $\Psi$ is a smooth function, it is quite straightforward.
Even if $\Psi$ is not a smooth function, such as an option payoff, one can obtain explicit expressions of $(V^{(0)},Z^{(0)})$
in terms of $X_t$, too. This is because,  one can derive an approximate joint transition density of general
diffusion processes by the asymptotic expansion~\footnote{We intend to use the result of
asymptotic expansion only for higher order approximations.}.
In the following, let us suppose that we have obtained the solutions up to a given order of asymptotic expansion,
and write each of them as a function of $x_t$:
\bea
\begin{cases}
& V_t^{(0)}=v^{(0)}(x_t) \\
& Z_t^{(0)}=z^{(0)}(x_t)
\end{cases}
\eea

\subsection{$\ep$-1st Order}
Since the BSDE is linear, we can integrate as before.
Here, let us first consider the evaluation of $V_t^{(1)}$.
\bea
V_t^{(1)}&=&\int_t^T \mathbb{E}\Bigl[f(X_u,V_u^{(0)},Z_u^{(0)})\Bigr|\calf_t\Bigr]du \nn\\
&=&\int_t^T \mathbb{E}\Bigl[f\Bigl(X_u,v^{(0)}(X_u),z^{(0)}(X_u)\Bigr)\Bigr|\calf_t\Bigr]du
\label{vt1}
\eea
Although it is possible to carry out standard Monte Carlo simulation for every time $u\in(t,T)$ and integrate 
to obtain the $V_t^{(1)}$, the time integration becomes numerically quite heavy. In fact,
it will soon become infeasible for $\ep$ higher order terms that include multi-dimensional integration 
of time.
We now introduce particle interpretation by McKean~\cite{McKean} developed for the study of semilinear PDEs:
\begin{proposition}
The $V_t^{(1)}$ in (\ref{vt1}) can be equivalently expressed as
\bea
V_t^{(1)}&=&\bold{1}_{\{\tau>t\}}\mathbb{E}\left[\left.\bold{1}_{\{\tau< T\}}
\hat{f}_t\Bigl(X_{\tau},v^{(0)}(X_{\tau}),z^{(0)}(X_{\tau})\Bigr)\right|\calf_t\right] 
\label{vt1_default}
\eea
Here $\tau$ is the time of interaction which is drawn independently from 
Poisson distribution with an arbitrary deterministic
positive intensity process $\lambda_t$. It can be a positive constant for 
the simplest case.~\footnote{It is not difficult to make it a stochastic process.}
$\hat{f}$ is defined as
\bea
\hat{f}_t(x,v^{(0)}(x),z^{(0)}(x)):=\frac{1}{\lambda_s}e^{\int_t^s \lambda_u du}f(x,v^{(0)}(x),z^{(0)}(x))~.
\eea
\label{prop1}
\end{proposition}
{\it Proof}: Define the new process for $(s>t)$:
\bea
\hat{V}_{t,s}^{(1)}=e^{\int_t^s \lambda_u du}V_s^{(1)}
\eea
then its dynamics is given by
\bea
d\hat{V}_{t,s}^{(1)}&=&e^{\int_t^s \lambda_u du}\left\{\lambda_s V_s^{(1)}ds
-f(X_s,v^{(0)}(X_s),z^{(0)}(X_s))ds+Z_s^{(1)}\cdot dW_s\right\} \nn \\
&=&\lambda_s \hat{V}_{t,s}^{(1)}ds-\lambda_s \hat{f}_{t}(X_s,v^{(0)}(X_s),z^{(0)}(X_s))ds
+e^{\int_t^s \lambda_u du}Z_s^{(1)}\cdot dW_s~.
\eea
Since we have $\hat{V}_{t,t}^{(1)}=V_t^{(1)}$, one can easily see the following relation holds:
\bea
&&V_t^{(1)}=\int_t^T \mathbb{E}\left[\left. e^{-\int_t^u \lambda_s ds}
\lambda_u \hat{f}_t(X_u,v^{(0)}(X_u),z^{(0)}(X_u))\right|\calf_t\right]du
\label{vt1_cva}
\eea
It is clear for those familiar with credit risk modeling~\cite{Bielecki, BJ}, it is nothing but the present value of 
default payment where the 
default intensity is $\lambda$ with the default payoff at $s~(>t)$ as $\hat{f}_t(X_s,v^{(0)}(X_s),z^{(0)}(X_s))$.
Thus, it is clear that (\ref{vt1_cva}) is equivalent to (\ref{vt1_default}). $\blacksquare$
\\

Now, let us consider the martingale component $Z^{(1)}$.
It can be expressed as
\bea
Z_t^{(1)}=\int_t^T \mathbb{E}\left[\left.\cald_t f\Bigl(X_u,v^{(0)}(X_u),z^{(0)}(X_u)\Bigr)\right|\calf_t\right]du
\label{zt1org}
\eea
We perform the similar transformation for $Z^{(1)}$ to make it easier to interpret in 
the interacting particle model.
Firstly, let us observe that the dynamics of Malliavin derivative of $V^{(1)}$ follows
\bea
d(\cald_t V_s^{(1)})&=&-(\cald_t X_s^i)\Bigl\{\part_i+\part_i v^{(0)}(X_s)\part_v+
\part_i z^{a(0)}(X_s)\part_{z^a}\Bigr\}f(X_s,v^{(0)}(X_s),z^{(0)}(X_s))ds \nn \\
&&+(\cald_t Z_s^{(1)})\cdot dW_s \\
\label{DVts1}
\cald_t V_t^{(1)}&=&Z_t^{(1)}
\eea
For lighten the notation, let us introduce a derivative operator
\bea
\nabla_i(x,v^{(0)},z^{(0)})=\part_i+\part_i v^{(0)}(x)\part_v+\part_i z^{a(0)}(x)\part_{z^a}
\eea
and  also 
\bea
f(x,v^{(0)},z^{(0)})\equiv f(x,v^{(0)}(x),z^{(0)}(x))
\eea
Now, we can write Eq.~(\ref{DVts1}) as
\bea
d(\cald_t V_s^{(1)})=-(\cald_t X_s^{i})\nabla_i(X_s,v^{(0)},z^{(0)})f(X_s,v^{(0)},z^{(0)})ds
+(\cald_t Z_s^{(1)})\cdot dW_s \nn
\eea

Define, for $(s>t)$, 
\bea
\widehat{\cald_t V_s^{(1)}}=e^{\int_t^s \lambda_u du}(\cald_t V_s^{(1)})
\eea
then its dynamics can be written as
\bea
&&d(\widehat{\cald_t V_s^{(1)}})=e^{\int_t^s \lambda_u du}\Bigl\{
\lambda_s (\cald_t V_s^{(1)})ds-(\cald_t X_s^i)\nabla_i(X_s,v^{(0)},z^{(0)})f(X_s,v^{(0)},z^{(0)})ds \nn \\
&&\hspace{25mm}+\cald_t Z_s^{(0)}\cdot dW_s\Bigr\} \nn \\
&&\qquad=\lambda_s(\widehat{\cald_t V_s^{(1)}})ds-\lambda_s(\cald_tX_s^i)\nabla_i(X_s,v^{(0)},z^{(0)})
\hat{f}_t(X_s,v^{(0)},z^{(0)})ds\nn \\
&&\qquad\quad+e^{\int_t^s \lambda_u du}(\cald_t Z_s^{(0)})\cdot dW_s
\eea
We have
\bea
\widehat{\cald_t V_t^{(1)}}=Z_t^{(1)}
\eea
and hence
\bea
Z_t^{(1)}=\int_t^T \mathbb{E}\left[\left.
e^{-\int_t^u \lambda_s ds}\lambda_s(\cald_t X_u^i)\nabla_i(X_u,v^{(0)},z^{(0)})\hat{f}_t(X_u,v^{(0)},z^{(0)})\right|
\calf_t\right]
\eea
\\
\\
Thus, following the same argument of the proposition~\ref{prop1}, we can conclude:
\begin{proposition}
$Z^{(1)}_t$ in (\ref{zt1org}) is equivalently expressed as
\bea
Z_t^{a(1)}=\bold{1}_{\{\tau>t\}}\mathbb{E}\left[\left. \bold{1}_{\{\tau<T\}}(Y_{t,\tau} \gamma(X_t))^i_a\nabla_i(X_\tau,v^{(0)},z^{(0)})
\hat{f}_t(X_\tau,v^{(0)},z^{(0)})\right|\calf_t\right]
\eea
where the definitions of random time $\tau$ and the intensity process $\lambda$ are the same 
as those in proposition~\ref{prop1}.
\end{proposition}
As we shall see later, interpreting $(X,Y)$ as a pair of particles allows an efficient Monte 
Carlo implementation. For the evaluation of $Z^{(1)}$ for example, one can consider it 
as an system of two particles $(X,Y)$, which have the intensity $\lambda$ 
of the interaction that produces 
\bea
(Y_{t\tau} \gamma(X_t))^i_a\nabla_i(X_\tau,v^{(0)},z^{(0)})
\hat{f}_t(X_\tau,v^{(0)},z^{(0)})
\eea
at the interaction point and annihilate altogether.
For $V^{(1)}$, the interpretation is much simpler. A single particle $X$ with
the decay rate of $\lambda$ leaves $\hat{f}$ at its decay point and vanishes.

\subsection{$\ep$-2nd Order}
For the $\ep$-2nd order, one can observe that
\bea
\label{vt2}
V_t^{(2)}&=&\int_t^T \mathbb{E}\left[\left.
\Bigl(V_u^{(1)}\part_v+Z_u^{a(1)}\part_{z^a}\Bigr)f(X_u,v^{(0)},z^{(0)})\right|\calf_t\right]du \\
\label{zt2}
Z_t^{(2)}&=&\int_t^T \mathbb{E}\left[\left.\cald_t\Bigl\{\Bigl(V_u^{(1)}\part_v+Z_u^{a(1)}\part_{z^a}\Bigr)f(X_u,v^{(0)},z^{(0)})\Bigr\}\right|\calf_t\right]du 
\eea
solve the BSDE (\ref{bsde2nd}). 
Its particle interpretation is available by the similar transformation.

Firstly, for $(s>t)$, let us define
\bea
\hat{V}_{t,s}^{(2)}=e^{\int_t^s \lambda_u du}V_s^{(2)}
\eea
with some appropriate intensity process $\lambda$.
Then it follows
\bea
d\hat{V}_{t,s}^{(2)}&=&\lambda_s \hat{V}_{t,s}^{(2)}ds-
\lambda_s (V_s^{(1)}\part_v+Z_s^{a(1)}\part_{z^a})\hat{f}_t(X_s,v^{(0)},z^{(0)})ds\nn \\
&&\quad+e^{\int_t^s \lambda_u du}Z_s^{(2)}\cdot dW_s
\eea
Observing that $\hat{V}_{t,t}^{(2)}=V_t^{(2)}$, one can confirm that
\bea
V_t^{(2)}=\bold{1}_{\{\tau_1>t\}}\mathbb{E}\left[\left.
\bold{1}_{\{\tau_1<T\}}\Bigl(V_{\tau_1}^{(1)}\part_v+Z_{\tau_1}^{a(1)}\part_{z^a}\Bigr)\hat{f}_t
(X_{\tau_1},v^{(0)},z^{(0)})\right|\calf_t\right]
\eea
where $\tau_1$ is the random interaction time with intensity $\lambda$.
Now, using the tower property of conditional expectations, one can conclude that
\begin{proposition}
$V_t^{(2)}$ in (\ref{vt2}) is equivalently expressed as
\bea
&&V_t^{(2)}=\bold{1}_{\{\tau_1>t\}}\mathbb{E}\Bigl[\bold{1}_{\{\tau_1<\tau_2<T\}}
(\part_v\hat{f}_{t,\tau_1})\hat{f}_{\tau_1,\tau_2}\Bigr|\calf_t\Bigr]\nn \\
&&\hspace{-10mm}+\bold{1}_{\{\tau_1>t\}}\mathbb{E}\Bigl[\bold{1}_{\{\tau_1<\tau_2<T\}}
\part_{z^a}\hat{f}_{t,\tau_1}\Bigl(Y_{\tau_1,\tau_2}\gamma_{\tau_1}\Bigr)^i_a
\nabla_{i,\tau_2}\hat{f}_{\tau_1,\tau_2}\Bigr|\calf_t\Bigr]
\eea
where we have defined 
\bea
&\hat{f}_{t,s}\equiv \hat{f}_t(X_s,v^{(0)}(X_s),z^{(0)}(X_s)) \nn \\
&\nabla_{i,s}\equiv\nabla_i(X_s,v^{(0)}(X_s),z^{(0)}(X_s))\nn \\
&\gamma_t\equiv\gamma(X_t)
\eea
and $\tau_1$ and $\tau_2$ are the two interaction times randomly drawn with intensity $\lambda$.
\end{proposition}

A particle interpretation for the first term is quite simple. A particle $X$ starts at $t$ follows
the diffusion (\ref{XSDE}) with (self) interaction intensity $\lambda$.
For the first interaction time $\tau_1$, it yields $\part_v\hat{f}_{t,\tau_1}$ and at the 2nd interaction
time $\tau_2$ it yields $\hat{f}_{\tau_1,\tau_2}$ and decays away. 
The expectation value can be evaluated by preparing a large number of particles $X$ starting from the same point
and obeying the same physical law but spend independent lives.
For the second term, the interpretation is more interesting.
A particle $X$ starts at time $t$ and follows the diffusion $(\ref{XSDE})$ with interaction intensity $\lambda$.
At the first interaction time $\tau_1$, it yields $\part_{z^a}\hat{f}_{t,\tau_1}$
and at the same time bears a new particle $Y$. After $\tau_1$, the two particles $(X,Y)$
follow the diffusions $(\ref{XSDE})$ and $(\ref{YSDE})$, respectively. They have interaction intensity $\lambda$,
and at the second interaction point $\tau_2$ they yield $(Y_{\tau_1,\tau_2}\gamma(X_{\tau_2}))
\nabla_{\tau_2}\hat{f}_{\tau_1,\tau_2}$ and annihilate altogether.
As in the first example, the expectation can be calculated by preparing a large number of particle $X$
at the same starting point.
\\
\\
{\it Remark: 
Note that, if we simply use Eqs.~(\ref{vt1}, \ref{vt2}) and the tower property, 
we have to handle a two-dimensional time integration. It makes naive implementation of Monte Carlo
simulation numerically too heavy. In our particle interpretation, 
this problem is solved by introducing random interaction times with some intensity $\lambda$.
One can choose appropriate size of intensity that produces enough amount of events 
for Monte Carlo simulation.}
\\
\\
We now consider an interacting particle interpretation of $Z^{(2)}$. 
For the evaluation of $Z^{(2)}$, we need to define the second order stochastic flow
for $(t<s<u)$:
\bea
(\Gamma_{t,s,u})^i_{jk}&=&\frac{\part^2}{\part x_t^j\part x_s^k} X_u^i=\frac{\part}{\part x_t^j}(Y_{s,u})^i_k
\eea
Since we have
\bea
(Y_{s,u})^i_k=\del^i_k+\int_s^u (Y_{s,v})^l_k(\part_l\gamma^i_0(X_v)dv+\part_l \gamma^i(X_v)\cdot dW_v)
\eea
it is easy to see that
\bea
(\Gamma_{t,s,u})^i_{j,k}&=&\int_s^u(\Gamma_{t,s,v})^l_{j,k}(\part_l\gamma^i_0(X_v)dv+\part_l \gamma^i(X_v)\cdot dW_v)\nn \\
&&+\int_s^u(Y_{t,v})^m_j(Y_{s,v})^l_k(\part_{lm}\gamma^i_0(X_v)dv+\part_{lm}\gamma^i(X_v)\cdot dW_v)
\eea
Note that we have $\Gamma_{t,s,s}=0$, regardless of time $s~(>t)$.
Using the second order stochastic flow, the Malliavin derivative of $Y$ can be written as
\bea
\cald^a_t(Y_{s,v})^i_k =(\Gamma_{t,s,v})^i_{j,k}(\gamma^j(X_t))_a=(\Gamma_{t,s,v}\gamma(X_t))^i_{k,a}
\eea

Consider the process of Malliavin derivative $\cald_t V_s^{(2)}$. One can write its dynamics
for $(t<s)$ as
\bea
&&d(\cald_t V_s^{(2)})=-\Bigl( (\cald_t V_s^{(1)})\part_v+(\cald_t Z_s^{a(1)})\part_{z^a}\Bigr)f(X_s,v^{(0)},z^{(0)})ds \nn \\
&&\qquad\quad-(\cald_t X_s^i)\Bigl\{V_s^{(1)}\nabla_{i,s}(\part_v f(X_s,v^{(0)},z^{(0)}))
+(Z_s^{a(1)})\nabla_{i,s}(\part_{z^a}f(X_s,v^{(0)},z^{(0)}))\Bigr\}ds\nn \\
&&\qquad\quad+\cald_t Z_s^{(2)}\cdot dW_s\\
&&\cald_t V_t^{(2)}=Z_t^{(2)}
\eea
As before, we define
\bea
\widehat{\cald_t V_s^{(2)}}=e^{\int_t^s \lambda_u du}(\cald_t V_s^{(2)})
\eea
then its dynamics satisfies the following SDE:
\bea
&&d(\widehat{\cald_t V_s^{(2)}})=\lambda_s (\widehat{\cald_t V_s^{(2)}})ds
-\lambda_s\Bigl[(\cald_t X_s^i)\Bigl(V_s^{(1)}\nabla_{i,s}(\part_v\hat{f}_{t,s})
+(Z_s^{a(1)})\nabla_{i,s}(\part_{z^a}\hat{f}_{t,s})\Bigr)\nn \\
&&\qquad+\Bigl( (\cald_t V_s^{(1)})\part_v+(\cald_t Z_s^{a(1)})\part_{z^a}\Bigr)\hat{f}_{t,s}\Bigr]ds
+e^{\int_t^s \lambda_u du}\cald_t Z_s^{(2)}\cdot dW_s \\
&&(\widehat{\cald_t V_t^{(2)}})=Z_t^{(2)}
\eea
Then, the same arguments leads to
\bea
Z_t^{(2)}&=&\bold{1}_{\{\tau_1>t\}}\mathbb{E}\left[
\bold{1}_{\{\tau_1<T\}}(\cald_t X_{\tau_1}^i)\Bigl(V_{\tau_1}^{(1)}\nabla_{i,\tau_1}(\part_v\hat{f}_{t,\tau_1})
+(Z_{\tau_1}^{a(1)})\nabla_{i,\tau_1}(\part_{z^a}\hat{f}_{t,\tau_1})\Bigr)\right.\nn \\
&&\left.\left.\hspace{15mm}+\bold{1}_{\{\tau_1<T\}}\Bigl(
(\cald_t V_{\tau_1}^{(1)})\part_v+(\cald_t Z_{\tau_1}^{a(1)})\part_{z^a}\Bigr)\hat{f}_{t,\tau_1}
\right|\calf_t\right]
\eea
using the random interaction time $\tau_1$.
\begin{proposition}
$Z_t^{(2)}$ in (\ref{zt2}) is equivalently expressed as
\bea
&&Z_t^{a(2)}=\bold{1}_{\{\tau_1>t\}}\mathbb{E}\Bigl[\bold{1}_{\{\tau_1<\tau_2<T\}}
(Y_{t,\tau_1}\gamma_t)^i_a\nabla_{i,\tau_1}(\part_v\hat{f}_{t,\tau_1})\hat{f}_{\tau_1,\tau_2} \Bigr|\calf_t\Bigr]\nn \\
&&+\bold{1}_{\{\tau_1>t\}}\mathbb{E}\Bigl[\bold{1}_{\{\tau_1<\tau_2<T\}}
(Y_{t,\tau_1}\gamma_t)^i_a\nabla_{i,\tau_1}(\part_{z^b}\hat{f}_{t,\tau_1})(Y_{\tau_1,\tau_2}\gamma_{\tau_1})^j_b
\nabla_{j,\tau_2}\hat{f}_{\tau_1,\tau_2}
\Bigr|\calf_t\Bigr] \nn \\
&&+\bold{1}_{\{\tau_1>t\}}\mathbb{E}\Bigl[\bold{1}_{\{\tau_1<\tau_2<T\}}
(\part_v \hat{f}_{t,\tau_1})(Y_{t,\tau_2}\gamma_t)^i_a\nabla_{i,\tau_2}\hat{f}_{\tau_1,\tau_2}\Bigr|\calf_t\Bigr] \nn \\
&&+\bold{1}_{\{\tau_1>t\}}\mathbb{E}\Bigl[\bold{1}_{\{\tau_1<\tau_2<T\}}
(\part_{z^b}\hat{f}_{t,\tau_1})(\gamma_{\tau_1})^j_b(\Gamma_{t,\tau_1,\tau_2}\gamma_t)^i_{j,a}\nabla_{i,\tau_2}\hat{f}_{\tau_1,\tau_2}
\Bigr|\calf_t\Bigr]\nn \\
&&+\bold{1}_{\{\tau_1>t\}}\mathbb{E}\Bigl[\bold{1}_{\{\tau_1<\tau_2<T\}}
(\part_{z^b}\hat{f}_{t,\tau_1})(Y_{t,\tau_1}\gamma_t)^j_a(\part_j \gamma_{\tau_1})^k_b
(Y_{\tau_1,\tau_2})^i_k\nabla_{i,\tau_2}\hat{f}_{\tau_1,\tau_2}
\Bigr|\calf_t\Bigr]\nn \\
&&+\bold{1}_{\{\tau_1>t\}}\mathbb{E}\Bigl[\bold{1}_{\{\tau_1<\tau_2<T\}}
(\part_{z^b}\hat{f}_{t,\tau_1})(Y_{t,\tau_2}\gamma_t)^j_a(Y_{\tau_1,\tau_2}\gamma_{\tau_1})^i_b
\nabla_{j,\tau_2}(\nabla_{i,\tau_2}\hat{f}_{\tau_1,\tau_2})\Bigr|\calf_t\Bigr]~.
\eea
where $\tau_1$ and $\tau_2$ are sequential interaction times with intensity $\lambda$.
\end{proposition}
{\it Proof:} It can be shown straightforwardly by using the tower property of conditional expectations
and commutativity between the indicator functions and the Malliavin derivative 
due to the independence of $\lambda$. $\blacksquare$

\begin{figure}[!htb]
\hspace{20mm}\includegraphics[width=100mm]{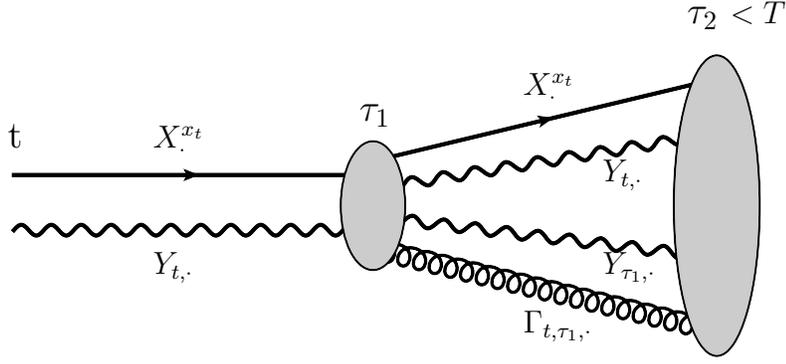}
\put(-285,80){\Large{t}}
\put(-230,80){\large $X_\cdot^{x_t}$}
\put(-230,32){\large $Y_{t,\cdot}$}
\put(-152,90){\Large $\tau_1$}
\put(-28,127){\Large $\tau_2$ \large $<T$}
\put(-90,100){\large $X^{x_t}_\cdot$}
\put(-60,67){\large $Y_{t,\cdot}$}
\put(-60,32){\large $Y_{\tau_1,\cdot}$}
\put(-90,10){\large $\Gamma_{t,\tau_1,\cdot}$}
\caption{A particle interpretation for $Z_t^{(2)}$.}
\label{graphz2}
\end{figure} 

Despite the apparent complexity, required numerical procedures for the evaluation of $Z^{(2)}$ 
is, in fact, quite simple. We provide a Feynman diagram for the particle interpretation 
in Figure~\ref{graphz2}. At the first stage, there are two particles of $(X_\cdot,Y_{t,\cdot})$
with initial values $(x_t, \{\del^i_j\})$, which survive until the second interaction time $\tau_2~(<T)$.
At the first interaction at $\tau_1$, two additional particles $(Y_{\tau_1,\cdot},\Gamma_{t,\tau_1,\cdot})$ 
are created. Each interaction occurs randomly with intensity $\lambda$.
Note that we already know the initial values of the new particles 
regardless of the interaction time, which makes numerical simulations possible to carry out.
What one has to do is to store the information of $\tau_1$ and $\tau_2$ and the values of 
the particles at these times. Then, all the ingredients in expectations can be calculated.
Simply repeating independent experiments and taking average will give the desired values.

\subsection{$\ep$-3rd Order: $V^{(3)}$}
In the similar fashion, we can proceed to higher order.
As before, by considering the dynamics of
\bea
\hat{V}_{t,s}^{(3)}=e^{\int_t^s \lambda_u du}V_s^{(3)}
\eea
one can observe that
\bea
&&V_t^{(3)}=\bold{1}_{\{\tau_1>t\}}\mathbb{E}\left[
\bold{1}_{\{\tau_1<T\}}\Bigl(
V_{\tau_1}^{(2)}\part_v+Z_{\tau_1}^{a(2)}\part_{z^a}+\frac{1}{2}(V_{\tau_1}^{(1)})^2\part_v^2 \right.\nn \\
&&\hspace{25mm}\left.\left.+V_{\tau_1}^{(1)}Z_{\tau_1}^{a(1)}\part_v\part_{z^a}+\frac{1}{2}Z_{\tau_1}^{a(1)}
Z_{\tau_1}^{b(1)}\part_{z^az^b}\Bigr)\hat{f}_{t,\tau_1}\right|\calf_t\right]~.
\eea
It can be written in terms of the fundamental variables simply applying tower property.

\begin{proposition}
$V_t^{(3)}$ can be expressed as
\bea
\label{v3first}
V_t^{(3)}&=&\bold{1}_{\{\tau_1>t\}}\mathbb{E}\Bigl[\bold{1}_{\{\tau_1<\tau_2<\tau_3\}}
(\part_v \hat{f}_{t,\tau_1})\Bigl\{(\part_v\hat{f}_{\tau_1,\tau_2})\hat{f}_{\tau_2,\tau_3}+(\part_{z^a}\hat{f}_{\tau_1,\tau_2})(Y_{\tau_2,\tau_3}\gamma_{\tau_2})^i_a
\nabla_{i,\tau_3}\hat{f}_{\tau_2,\tau_3}\Bigr\}\nn \\
&&+\bold{1}_{\{\tau_1<\tau_2<\tau_3\}}(\part_{z^a}\hat{f}_{t,\tau_1})\Bigl\{
(Y_{\tau_1,\tau_2}\gamma_{\tau_1})^i_a\nabla_{i,\tau_2}(\part_v\hat{f}_{\tau_1,\tau_2})\hat{f}_{\tau_2,\tau_3}\nn \\
&&+(Y_{\tau_1,\tau_2}\gamma_{\tau_1})^i_a\nabla_{i,\tau_2}(\part_{z^b}\hat{f}_{\tau_1,\tau_2})
(Y_{\tau_2,\tau_3}\gamma_{\tau_2})^j_b\nabla_{j,\tau_3}\hat{f}_{\tau_2,\tau_3}\nn \\
&&+(\part_{z^b}\hat{f}_{\tau_1,\tau_2})(\gamma_{\tau_2})^j_b
(\Gamma_{\tau_1,\tau_2,\tau_3}\gamma_{\tau_1})^i_{j,a}\nabla_{i,\tau_3}\hat{f}_{\tau_2,\tau_3}\nn \\
&&+(\part_{z^b}\hat{f}_{\tau_1,\tau_2})(Y_{\tau_1,\tau_2}\gamma_{\tau_1})^j_a
(\part_j\gamma_{\tau_2})^k_b(Y_{\tau_2,\tau_3})^i_k\nabla_{i,\tau_3}\hat{f}_{\tau_2,\tau_3}\nn \\
&&+(\part_{z^b}\hat{f}_{\tau_1,\tau_2})(Y_{\tau_1,\tau_3}\gamma_{\tau_1})^j_a(Y_{\tau_2,\tau_3}\gamma_{\tau_2})^i_b
\nabla_{j,\tau_3}(\nabla_{i,\tau_3}\hat{f}_{\tau_2,\tau_3})\Bigr\}\Bigr|\calf_t\Bigr]\\
\label{v3second}
&+&\bold{1}_{\{\tau_1>t\}}\mathbb{E}\left[\bold{1}_{\{\tau_1<T\}}
\frac{1}{2}(\part_v^2\hat{f}_{t,\tau_1})\prod_{p=1}^2\Bigl(
\bold{1}_{\{\tau_1<\tau_2^p<T\}}\hat{f}_{\tau_1,\tau_2^p}\Bigr) \right.\nn \\
&&+\bold{1}_{\{\tau_1<T\}}(\part_v\part_{z^a}\hat{f}_{t,\tau_1})
\Bigl(\bold{1}_{\{\tau_1<\tau_2^p<T\}}\hat{f}_{\tau_1,\tau_2^p}\Bigr)^{p=1}
\Bigl(\bold{1}_{\{\tau_1<\tau_2^p<T\}}(Y_{\tau_1,\tau_2^p}\gamma_{\tau_1})^i_a\nabla_{i,\tau_2^p}\hat{f}_{\tau_1,\tau_2^p}
\Bigr)^{p=2}\nn\\
&&+\bold{1}_{\{\tau_1<T\}}\frac{1}{2}(\part_{z^az^b}\hat{f}_{t,\tau_1})
\Bigl(\bold{1}_{\{\tau_1<\tau_2^p<T\}}(Y_{\tau_1,\tau_2^p}\gamma_{\tau_1})^i_a\nabla_{i,\tau_2^p}\hat{f}_{\tau_1,\tau_2^p}
\Bigr)^{p=1}\nn\\
&&\left.\left.\times\Bigl(\bold{1}_{\{\tau_1<\tau_2^p<T\}}(Y_{\tau_1,\tau_2^p}\gamma_{\tau_1})^j_b
\nabla_{j,\tau_2^p}\hat{f}_{\tau_1,\tau_2^p}\Bigr)^{p=2}\hspace{5mm}\right|\calf_t\right]
\eea
where the contents within each bracket of $p\in\{1,2\}$ must be calculated according to the diffusion processes 
$(X_\cdot^{x_{\tau_1}}, Y_{\tau_1,\cdot})_{p=\{1,2\}}$ that follow the identical diffusion laws with the same initial values,
but are independent with each other. $\{\tau_i\}_{i\geq 1}$ are sequential random times of interactions 
drawn with intensity $\lambda$. $\{\tau_2^p\}_{p=1,2}$ should be drawn independently.
\end{proposition}

\begin{figure}[!htb]
\hspace{25mm}\includegraphics[width=95mm]{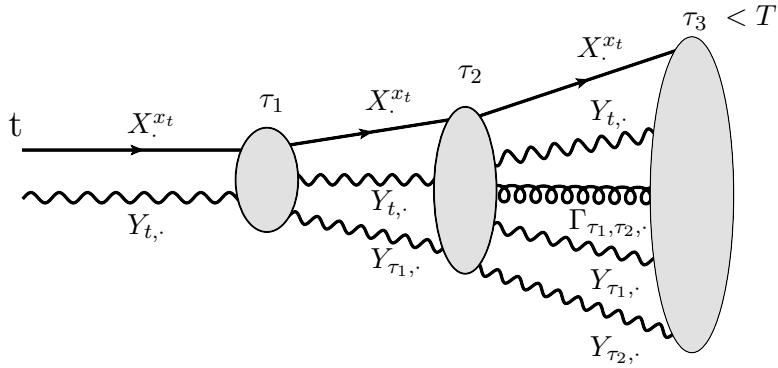}
\put(-275,83){\Large{t}}
\put(-230,84){$X_\cdot^{x_t}$}
\put(-230,46){$Y_{t,\cdot}$}
\put(-180,94){$\tau_1$}
\put(-140,93){$X^{x_t}_\cdot$}
\put(-60,113){$X^{x_t}_\cdot$}
\put(-55,89){$Y_{t,\cdot}$}
\put(-138,55){$Y_{t,\cdot}$}
\put(-138,33){$Y_{\tau_1,\cdot}$}
\put(-105,105){$\tau_2$}
\put(-63,48){$\Gamma_{\tau_1,\tau_2,\cdot}$}
\put(-55,25){$Y_{\tau_1,\cdot}$}
\put(-55,0){$Y_{\tau_2,\cdot}$}
\put(-20,125){$\tau_3~<T$}
\caption{A particle interpretation for the first half of $V_t^{(3)}$.}
\label{graphV3_first}
\end{figure}

\begin{figure}[!htb]
\hspace{35mm}\includegraphics[width=90mm]{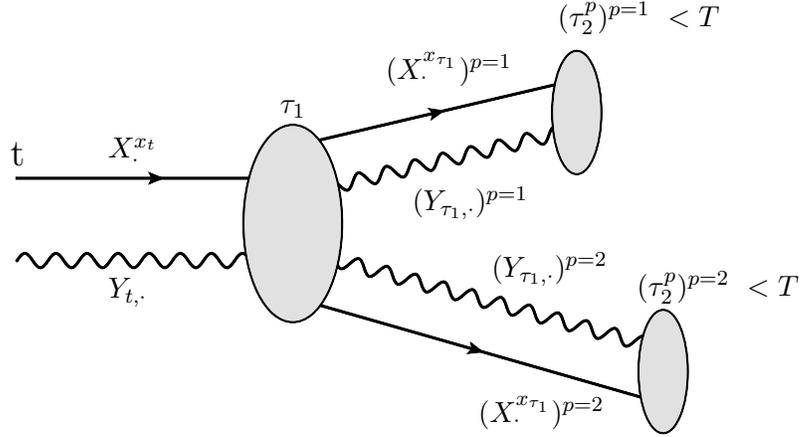}
\put(-257,102){\Large{t}}
\put(-220,105){$X_\cdot^{x_t}$}
\put(-220,52){$Y_{t,\cdot}$}
\put(-155,122){$\tau_1$}
\put(-115,135){$(X^{x_{\tau_1}}_\cdot)^{p=1}$}
\put(-80,5){$(X^{x_{\tau_1}}_\cdot)^{p=2}$}
\put(-105,85){$(Y_{\tau_1,\cdot})^{p=1}$}
\put(-75,60){$(Y_{\tau_1,\cdot})^{p=2}$}
\put(-50,155){$(\tau_2^p)^{p=1}~<T$}
\put(-20,53){$(\tau_2^p)^{p=2}~<T$}
\caption{A particle interpretation for the second half of $V_t^{(3)}$.}
\label{graphV3_2nd}
\end{figure}

Note that, we have introduced two sets of particles labeled by $p\in\{1,2\}$ that follow the same physical laws
but perfectly independent with each other to eliminate $\tau_1$-conditional expectations. 
In this way, one can avoid the use of nested Monte Carlo simulation.
In Figures~\ref{graphV3_first} and \ref{graphV3_2nd}, we have provided two Feynman diagrams, one for the first half,
and the other for the second half of the expression of $V_t^{(3)}$.
In simulations, one has to store the interaction times and 
all the relevant particles values at those points to evaluate the expectations.

\subsection{$Z^{(3)}$ and $\ep$-higher order terms}
The valuation procedures for $Z^{(3)}$ are almost the same as that of $Z^{(2)}$, but we need to introduce
a new type of particle corresponding to $(\frac{\part}{\part x_t}\Gamma_{s,u,v,\cdot})$.
As easily guessed from the previous examples,
we need to add one new particle corresponding to a higher order stochastic flow
to complete the particle picture at every time when we proceed a $\ep$-higher order
approximation (of martingale component $Z$). A remarkable fact is that 
all the initial conditions of the new particles created at random times
are known beforehand thanks to the characteristics of the Malliavin derivatives.
This feature makes one can perform numerical simulations that describe full history of 
the evolution of particles.

\section{Extension to Fully-Coupled Cases}
We now consider the situation where the underlying state process $X$ also gets
the feedbacks from the backward components.  By making use of the perturbative technique
in PDE framework~\cite{analytic_FBSDE}, we shall show that 
the same strategy in the previous sections works well also in this seemingly much 
more complicated situation.

The dynamics of whole system is given by
\bea
\begin{cases}
&dV_t=-f(t,X_t,V_t,Z_t)dt+Z_t\cdot dW_t \\
& V_T=\Psi(X_T) \\
&dX_t=\gamma_0(t,X_t,V_t,Z_t)dt+\gamma(t,X_t,V_t,Z_t)\cdot dW_t \\
&X_0=x
\end{cases}
\eea
where we have distinguished time arguments from $X$ to make PDE generator a familiar form.
As before, we assume that $V,Z,X$ take value in $\mathbb{R}, \mathbb{R}^r$ and
$\mathbb{R}^d$ respectively, and $W$ denotes a $r$-dimensional Brownian motion.
Following the idea of four-step scheme~\cite{fourstep}, we postulate
that $V_t$ is given by some appropriate function of $t$ and $X$, $v(t,X)$.
Then it needs to satisfy the relevant PDE:
\bea
\begin{cases}
&\part_t v(t,x)+\Bigl\{\part_i v(t,x)\gamma^i_0(t,x,v(t,x),z(t,x))+
\frac{1}{2}\part_{ij}v(t,x)(\gamma^i\cdot \gamma^j)(t,x,v(t,x),z(t,x))\Bigr\}  \\
&\qquad\qquad+f(t,x,v(t,x),z(t,x))=0\\
&z(t,x)=\part_iv(t,x)\gamma^i(t,x,v(t,x),z(t,x)) \\
&v(T,x)=\Psi(T,x)
\end{cases}
\eea

The above non-linear PDE cannot be solved in general. Therefore, let us introduce perturbation parameter $\ep$ as before,
\bea
\begin{cases}
&dV_t^{(\ep)}=-\ep f(t,X_t^{(\ep)},V_t^{(\ep)},Z_t^{(\ep)})dt+Z_t^{(\ep)}\cdot dW_t \\	
&V_T^{(\ep)}=\Psi(X_T^{(\ep)})\\
&dX_t^{(\ep)}=\Bigl(r(t,X_t^{(\ep)})+\ep\mu(t,X_t^{(\ep)},V_t^{(\ep)},Z_t^{(\ep)})\Bigr)dt\nn \\
&\hspace{35mm} +\Bigl(\sigma(t,X_t^{(\ep)})+\ep\eta(t,X_t^{(\ep)},V_t^{(\ep)},Z_t^{(\ep)})\Bigr)\cdot dW_t\\
&X_0^{(\ep)}=x
\end{cases}
\eea
and its corresponding PDE
\bea
\begin{cases}
&\part_i v^{(\ep)}(t,x)+\Bigl\{\part_i v^{(\ep)}(t,x)\gamma_0^i(t,x,v^{(\ep)},z^{(\ep)})+
\frac{1}{2}\part_{ij}v^{(\ep)}(t,x)(\gamma^i\cdot\gamma^j)(t,x,v^{(\ep)},z^{(\ep)})\Bigr\} \\
&\qquad+\ep f(t,x,v^{(\ep)},z^{(\ep)})=0\\
&z^{(\ep)}(t,x)=\part_i v^{(\ep)}(t,x)\gamma^i(t,x,v^{(\ep)},z^{(\ep)})\\
&v^{(\ep)}(T,x)=\Psi(x)
\label{PDE_perturbative}
\end{cases}
\eea
Here, we have extracted the terms free from feedback effects from $X$'s 
dynamics~\footnote{Although this can be done somewhat arbitrarily, it may be natural to set $r(t,x)$ and $\sigma(t,x)$ 
as the expected dynamics of $X$ when all the feedback effects are switched off.}:
\be
\begin{cases}
&\gamma_0(t,x,v^{(\ep)},z^{(\ep)})=r(t,x)+\ep \mu(t,x,v^{(\ep)}(t,x),z^{(\ep)}(t,x)) \\
&\gamma(t,x,v^{(\ep)},z^{(\ep)})=\sigma(t,x)+\ep\eta(t,x,v^{(\ep)}(t,x),z^{(\ep)}(t,x))
\end{cases}
\label{Xperturb}
\ee

We suppose that the solution of the above PDE can be expanded perturbatively in terms of $\ep$ as
\bea
&v^{(\ep)}(t,x)=v^{(0)}(t,x)+\ep v^{(1)}(t,x)+\ep^2 v^{(2)}(t,x)+\cdots \\
&z^{(\ep)}(t,x)=z^{(0)}(t,x)+\ep z^{(1)}(t,x)+\ep^2 z^{(2)}(t,x)+\cdots 
\eea
As in the previous sections, putting $\ep=1$ is expected to give the approximation of the 
original system as long as the non-linear effects are perturbative.

\subsection{Expansion of non-linear PDE}
Straightforward calculation allows us to expand the original PDE into a 
series of linear parabolic PDEs. See \cite{analytic_FBSDE} for details.
Firstly, let us define the differential operator $\call$:
\bea
\call(t,x)=r^i(t,x)\part_i+\frac{1}{2}(\sigma^i\cdot \sigma^j)(t,x)\part_{ij}
\eea
which corresponds to the infinitesimal generator of $X^{(0)}$, ie., the free forward component
\bea
dX_t^{(0)}&=&r(t,X_t)dt+\sigma(t,X_t)\cdot dW_t \\
X_0^{(0)}&=&x
\eea

Using this generator, we can show that the backward components in each order satisfy:\\
{\bf{Zero-th order}}
\bea
\begin{cases}
 &\Bigl(\part_t+\call(t,x)\Bigr)v^{(0)}(t,x)=0\\
 &v^{(0)}(T,x)=\Psi(x)
\end{cases}
\eea
and
\be
 \qquad z^{(0)}(t,x)=\part_i v^{(0)}(t,x)\sigma^i(t,x)
\ee
{{\bf{Higher expansion order $(n\geq 1)$}}
\bea
\begin{cases}
&\Bigl(\part_t+\call(t,x)\Bigr)v^{(n)}(t,x)+G^{(n)}(t,x)=0 \\
&v^{(n)}(T,x)=0
\end{cases}
\label{higher_PDE}
\eea
where the expression of $G^{(n)}$ and $z^{(n)}$ can be obtained straightforwardly
by extracting $\calo(\ep^n)$ terms from (\ref{PDE_perturbative}).

\subsection{Particle Interpretation}
The crucial point in the previous subsection is, because of the perturbation structure in (\ref{Xperturb}),
the relevant differential operator always derived from $X^{(0)}$ and remains the same
for all the expansion orders. In addition, since we put a $\ep$-factor in front of the non-linear terms,
$G^{(n)}$ contains the backward components with $\ep$-order only up to $(n-1)$.
Furthermore, it is clear to see that $z^{(n)}$ can only contain the backward components of $\{v^{(m)}\}_{\{m\leq n\}}$
and $\{z^{(m)}\}_{\{m\leq (n-1)\}}$.
Therefore, using Feynman-Kac theorem, we see that the PDE in (\ref{higher_PDE}) is equivalently expressed by
\bea
\begin{cases}
&dV_t^{(n)}=-G^{(n)}(t,X_t^{(0)},V^{(n-1)}_t,Z^{(n-1)}_t,\cdots)dt+\tilde{Z}^{(n)}_t\cdot dW_t \\
&V_T^{(n)}=0
\end{cases}
\eea
where the dynamics of the forward component $X^{(0)}$ is already known.
Because of the very nature of the perturbative expansion, 
all the $(V^{(m)},Z^{(m)})_{\{m\geq 1\}}$ appear as a power series and not contained within 
the non-linear functions.
Thus, $V_t^{(n)}$ can be solved by the same procedures studied in the previous sections,
and also the nice properties of explicitly capped number of branches and interaction points
still hold.

Note that $\tilde{Z}_t^{(n)}$ is not equal to $Z_t^{(n)}$ that contains additional terms 
through the feedbacks to $X$. However, it is not difficult to calculate these terms.
For example, one can observe:\\
{\bf{1st order $(n=1)$}}
\bea
G^{(1)}(t,x)&=&f^{(0)}(t,x) +\part_i v^{(0)}(t,x)\mu^{i(0)}(t,x)+\part_{ij}v^{(0)}(t,x)(\sigma^i\cdot\eta^{j(0)})(t,x) \\
z^{(1)}(t,x)&=&\part_i v^{(1)}(t,x)\sigma^i(t,x)+\part_iv^{(0)}(t,x)\eta^{i(0)}(t,x)
\label{z1pde}
\eea
{\bf{2nd order $(n=2)$}}
\bea
G^{(2)}(t,x)&=&\bigl(v^{(1)}(t,x)\part_v+z^{a(1)}(t,x)\part_{z^a}\bigr)f^{(0)}(t,x)\nn \\
&&+\part_i v^{(1)}(t,x)\mu^{i(0)}(t,x)+\part_i v^{(0)}(t,x)
\bigl(v^{(1)}(t,x)\part_v+z^{a(1)}(t,x)\part_{z^a}\bigr)\mu^{i(0)}(t,x)\nn \\
&&+\part_{ij}v^{(1)}(t,x)(\sigma^i\cdot\eta^{j(0)})(t,x)+\frac{1}{2}\part_{ij}v^{(0)}(t,x)
(\eta^{i(0)}\cdot\eta^{j(0)})(t,x)\nn \\
&&+\part_{ij}v^{(0)}(t,x)\sigma^i(t,x)\cdot \bigl(v^{(1)}(t,x)\part_v+z^{a(1)}(t,x)\part_{z^a}\bigr)\eta^{j(0)}(t,x)\\
z^{(2)}(t,x)&=&\part_i v^{(2)}(t,x)\sigma^i(t,x)+\part_i v^{(1)}(t,x)\eta^{i(0)}(t,x)\nn\\
&&+\part_iv^{(0)}(t,x)\bigl(v^{(1)}(t,x)\part_v+z^{a(1)}(t,x)\part_{z^a}\bigr)\eta^{i(0)}(t,x)
\eea
Higher order cases can be obtained similarly.

Let us now consider the particle method to evaluate the relevant terms. Let us fix
the initial time as $t$ as before: For the zero-th order, the problem is exactly the same as the decoupled case and 
we can derive easily $v^{(0)}(t,x)$ and $z^{(0)}(t,x)$ as a function of $x$ by asymptotic 
expansion~\footnote{As before, this is only to use higher order expansion. For the valuation of the 
zero-th order itself, one can use the standard Monte Carlo simulation}. 
For simplicity, we write $X_s^{(0)}$ as $X_s$, since the underlying process does not change. 
\\
\\
{\bf{1st order}}\\
As for the first order, observe that $G^{(1)}(t,x)$ is given as an explicit function of $x$
after the completion of the zero-th order calculation.
Then, $V_s^{(1)}$ follows
\bea
\begin{cases}
&dV^{(1)}_s=-G^{(1)}(t,X_s)ds+\tilde{Z}_s^{(1)}\cdot dW_s  \\
&V_T^{(1)}=0
\end{cases}
\eea
and hence, by the same arguments, for $(s>t)$, we have a particle representation as
\bea
V_t^{(1)}=\bold{1}_{\{\tau>t\}}\mathbb{E}\Bigl[\bold{1}_{\{\tau<T\}}\hat{G}^{(1)}_{t}(\tau,X_{\tau})\Bigr|\calf_t\Bigr]
\eea
where $\hat{G}_t^{(1)}$ is defined as
\bea
\hat{G}^{(1)}_t(s,X_{s})=\frac{1}{\lambda_s}e^{\int_t^s \lambda_u du}G^{(1)}(s,X_s)
\eea
with some appropriate positive deterministic (or independent) intensity $\lambda$.
For martingale component, it is easy to see
\bea
Z_t^{(1)}=\tilde{Z}^{(1)}_t+\part_i v^{(0)}(t,x_t)\eta^{i(0)}(t,x_t)
\eea
from (\ref{z1pde}). Here, the particle representation of 
$\tilde{Z}^{(1)}$ can be derived in the same way as in the decoupled case:
\bea
\tilde{Z}_t^{a(1)}=\bold{1}_{\{\tau>t\}}\mathbb{E}\Bigl[
\bold{1}_{\{\tau<T\}}(Y_{t,\tau}\sigma_t)^i_a\part_i\hat{G}_t^{(1)}(\tau,X_{\tau})\Bigr|\calf_t\Bigr]
\eea
where $Y_{t,s}~(s>t)$ is the stochastic flow of $X$ and is given by
\bea
(Y_{t,u})^i_j=\del_j^i+\int_t^u (Y_{t,s})^k_j\Bigl\{
\part_kr^i(s,X_s)ds+\part_k \sigma^i(s,X_s)\cdot dW_s\Bigr\}
\eea
The second term of $Z^{(1)}$ is already 
given as an explicit function of $x_t$. 
\\
\\
{\bf{2nd order}}\\
We can proceed to higher orders in similar fashion.
For the second order, the contribution to $V^{(2)}$ from the first line of $G^{(2)}$ can be calculated 
in the same way as the decoupled case. Let us consider non-trivial remaining terms.
The contribution from $\part_i v^{(1)}(t,x)\mu^{i(0)}(t,x)$, for example, can be calculated as
\bea
&&\bold{1}_{\{\tau_1>t\}}\mathbb{E}\left[\left.\bold{1}_{\{\tau_1<T\}}\hat{\mu}_t^{i(0)}(\tau_1,X_{\tau_1})
\frac{\part}{\part x_{\tau_1}^i}\Bigl(\bold{1}_{\{\tau_2>\tau_1\}}
\mathbb{E}\Bigl[ \bold{1}_{\{\tau_2<T\}}\hat{G}^{(1)}_{\tau_1}(\tau_2,X_{\tau_2})\Bigr|\calf_{\tau_1}\Bigr]\Bigr)
\right|\calf_t\right]\nn \\
&&=\bold{1}_{\{\tau_1>t\}}\mathbb{E}\left[\left.
\bold{1}_{\{\tau_1<\tau_2<T\}}\hat{\mu}_t^{i(0)}(\tau_1,X_{\tau_1})
(Y_{\tau_1,\tau_2})^j_i\part_j\hat{G}^{(1)}_{\tau_1}(\tau_2,X_{\tau_2})\right|\calf_t\right]
\eea
where 
\be
\hat{\mu}_t^{i(0)}(s,X_s)=\frac{1}{\lambda_s}e^{\int_t^s \lambda_u du}\mu^{i(0)}(s,X_s)
\ee
Note that the partial derivative of $x$ in $\part_i v^{(1)}(\tau_1,X_{\tau_1})$ should be recognized
as the shift of $X$ at the time of $\tau_1$, which leads to the first order stochastic flow $Y$ in the
above expression. 

Next, let us consider the contribution from $\part_{ij}v^{(1)}(t,x)(\sigma^i\cdot\eta^{j(0)})(t,x)$.
As is the previous example, it is calculated as
\bea
&&\bold{1}_{\{\tau_1>t\}}\mathbb{E}\left[\left.\bold{1}_{\{\tau_1<T\}}(\widehat{\sigma^i\cdot\eta^{j(0)}})(\tau_1,X_{\tau_1})
\frac{\part^2}{\part x^i_{\tau_1}\part x^j_{\tau_1}}\Bigl(
\bold{1}_{\{\tau_2>\tau_1\}}\mathbb{E}\Bigl[\bold{1}_{\{\tau_2<T\}}\hat{G}^{(1)}_{\tau_1}(\tau_2,X_{\tau_2})\Bigr|
\calf_{\tau_1}\Bigr]
\Bigr)\right|\calf_t\right] \nn \\
&&=\bold{1}_{\{\tau_1>t\}}\mathbb{E}\left[\bold{1}_{\{\tau_1<\tau_2<T\}}
(\widehat{\sigma^i\cdot\eta^{j(0)}})(\tau_1,X_{\tau_1})\Bigl\{
(\Gamma_{\tau_1,\tau_2})^k_{ij}\part_k\hat{G}^{(1)}_{\tau_1}(\tau_2,X_{\tau_2})\right.
\nn \\
&&\hspace{65mm}+(Y_{\tau_1,\tau_2})^k_i(Y_{\tau_1,\tau_2})^l_j\part_{kl}\hat{G}^{(1)}_{\tau_1}(\tau_2,X_{\tau_2})
\Bigr\}\Bigr|\calf_t\Bigr]
\eea
where $\widehat{\sigma^i\cdot\eta^{j(0)}}$ is defined similarly as $\hat{G}^{(1)}$. Note that
the second order stochastic flow $(\Gamma_{t,s})^k_{i,j}$
is defined, for $(u>t)$, as
\bea
(\Gamma_{t,u})^k_{i,j}=\frac{\part}{\part x_t^i\part x_t^j}(X_u^{x_t})^{k}
\eea
and is given by
\bea
(\Gamma_{t,u})^k_{i,j}&=&\int_t^u (\Gamma_{t,s})^l_{ij}\Bigl\{
\part_l r^k(s,X_s)ds+\part_l \sigma^k(s,X_s)\cdot dW_s\Bigr\} \nn \\
&&+\int_t^u (Y_{t,s})^l_i(Y_{t,s})^m_j\Bigl\{
\part_{lm}r^k(s,X_s)ds+\part_{lm}\sigma^k(s,X_s)\cdot dW_s\Bigr\}
\eea
The remaining contributions to $V^{(2)}$ as well as $Z^{(2)}$ can be calculated by 
the same technique. 
\begin{center}
$\cdots\cdots\cdots$
\end{center}
Although tedious calculation is needed, we can proceed to an arbitrary higher order
in the same fashion. Note that, also in fully-coupled cases, new particles required in simulation 
are all derived as stochastic flows of $X$ and hence the initial values at their creations are known beforehand.
 
\section{Conclusions and Discussions}
In this paper, we have developed an efficient Monte Carlo scheme 
with an interacting particle representation. It allows 
straightforward numerical implementation to solve fully non-linear decoupled as well as coupled FBSDEs
at each order of perturbative expansion.
The appearance of unknown backward components in the expressions of higher order approximations
is solved by introducing an appropriate particle interpretation.
Although a couple of new particles are created at random interaction times,
their initial values are known beforehand. This is due to their properties
as the stochastic flows of the underlying sate, which is the crucial point 
to make straightforward Monte Carlo simulation possible.
The proposed method can be applied to semi-linear problems, such as American and Bermudan options,
Credit Value Adjustment (CVA), and even fully non-linear issues, such as the optimal portfolio problems in 
incomplete and/or constrained markets, feedbacks from large investors,  
and also the analysis of various risk measures.
It looks also interesting to use the current method
to study higher order FBSDEs, where the higher order Malliavin derivatives
exist in the non-linear driver, such as $f(t,X_t,V_t,\cald_t V,\cald_t^2 V)$.
It can be done straightforwardly by introducing higher order stochastic flows.
\\
\\
{\bf Acknowledgment:} The authors thank Seisho Sato of the Institute of Statistical Mathematics (ISM) 
for the helpful discussions about the branching diffusion method.


\end{document}